\documentclass[prl,twocolumn,showpacs,preprintnumbers,nofootinbib]{revtex4-1}
\usepackage{amsmath,bm,graphicx,hyperref}

\newcommand\<{\langle}
\renewcommand\>{\rangle}
\newcommand\e{{\hat{\bm{e}}}}
\renewcommand\r{{\bm{r}}}
\newcommand\rh{\bm\rho}
\newcommand\R{\bm{R}}
\newcommand\x{{\bm{x}}}
\newcommand\y{{\bm{y}}}
\newcommand\z{{\bm{z}}}
\renewcommand\k{\bm{k}}
\newcommand\K{{\bm{K}}}
\renewcommand\S{{\bm{S}}}
\newcommand\0{{\bm{0}}}

\begin{document}
\preprint{LA-UR-12-24050}

\title{Efimov effect in quantum magnets}

\author{Yusuke~Nishida}\email{nishida@yukawa.kyoto-u.ac.jp}
\author{Yasuyuki~Kato}
\author{Cristian~D.~Batista}
\affiliation{Theoretical Division, Los Alamos National Laboratory,
Los Alamos, New Mexico 87545, USA}

\date{August 2012}

\begin{abstract}
 Physics is said to be universal when it emerges regardless of the
 underlying microscopic details.  A prominent example is the Efimov
 effect, which predicts the emergence of an infinite tower of three-body
 bound states obeying discrete scale invariance when the particles
 interact resonantly.  Because of its universality and peculiarity, the
 Efimov effect has been the subject of extensive research in chemical,
 atomic, nuclear and particle physics for decades.  Here we employ an
 anisotropic Heisenberg model to show that collective excitations in
 quantum magnets (magnons) also exhibit the Efimov effect.  We locate
 anisotropy-induced two-magnon resonances, compute binding energies of
 three magnons and find that they fit into the universal scaling law.
 We propose several approaches to experimentally realize the Efimov
 effect in quantum magnets, where the emergent Efimov states of magnons
 can be observed with commonly used spectroscopic measurements.  Our
 study thus opens up new avenues for universal few-body physics in
 condensed matter systems.
\end{abstract}


\maketitle

\subsection{Introduction}
Sometimes we observe that completely different systems exhibit the same
physics.  Such physics is said to be universal and its most famous
example is the critical phenomena~\cite{Fisher:1998}.  In the vicinity
of second-order phase transitions where the correlation length diverges,
microscopic details become unimportant and the critical phenomena are
characterized by only a few ingredients; dimensionality, interaction
range and symmetry of the order parameter.  Accordingly, fluids and
magnets exhibit the same critical exponents.  The universality in
critical phenomena has been one of the central themes in condensed
matter physics.

Similarly, we can also observe universal physics in the vicinity of
scattering resonances where the $s$-wave scattering length diverges.
Here low-energy physics is characterized solely by the $s$-wave
scattering length and does not depend on other microscopic details.  One
of the most prominent phenomena in such universal systems is the Efimov
effect, which predicts the emergence of an infinite tower of three-body
bound states obeying discrete scale invariance:
\begin{align}\label{eq:scaling}
 \frac{E_{n+1}}{E_n} \to \lambda^{-2} \qquad (n\to\infty)
\end{align}
with the universal scale factor
$\lambda=22.6944$~\cite{Efimov:1970}.  Because of its universality and
peculiarity, the Efimov effect has been the subject of extensive
research in chemical, atomic, nuclear and particle physics for decades
after the first prediction in 1970~\cite{Nielsen:2001,Braaten:2006}.  In
particular, the recent experimental realization with ultracold atoms has
greatly stimulated this research area~\cite{Kraemer:2006}.

In spite of such active research, the Efimov effect has not attracted
much attention in condensed matter physics so far.  However, because the
Efimov effect is universal, it should emerge also in condensed matter
systems.  In this article, we show that collective excitations in
quantum magnets (magnons) exhibit the Efimov effect by tuning an
easy-axis exchange or single-ion anisotropy.  We will locate
anisotropy-induced two-magnon resonances, compute the binding energies
of three magnons and find that they fit into the universal scaling law.
We will also propose several approaches to experimentally realize the
Efimov effect in quantum magnets, including frustrated cases.  So far
multi-magnon bound states have been observed with different experimental
techniques, but mostly in quasi-one-dimensional
compounds~\cite{Date:1966,Date:1968,Torrance:1969th,Torrance:1969ex,Nicoli:1974,Vlimmeren:1979,Vlimmeren:1980,Hoogerbeets:1984,Nijhof:1986,Bosch:1987,Rubins:1994,Perkins:1995,Garrett:1997,Orendac:1999,Katsumata:2000,Tennant:2003,Zvyagin:2007,Psaroudaki:2012}.
Although the Efimov effect emerges only in three
dimensions~\cite{Nielsen:2001,Braaten:2006}, the same spectroscopic
measurements can be used to observe the emergent Efimov states of
magnons.  Our study thus opens up new avenues for universal few-body
physics in condensed matter systems.  Also, in addition to the
Bose--Einstein condensation of magnons~\cite{Giamarchi:2008}, the Efimov
effect provides a novel connection between atomic and magnetic systems.

\subsection{Anisotropic Heisenberg model}
To demonstrate the Efimov effect in quantum magnets, we consider an
anisotropic Heisenberg model on a simple cubic lattice:
\begin{align}\label{eq:hamiltonian}
 H &= - \frac12\sum_\r\sum_\e
 (J\,S_\r^+S_{\r+\e}^- + J_z\,S_\r^zS_{\r+\e}^z) \notag\\
 &\quad - D\sum_\r(S_\r^z)^2 - B\sum_\r S_\r^z,
\end{align}
where $\sum_\e$ is a sum over six unit vectors;
$\sum_{\e=\pm\hat\x,\pm\hat\y,\pm\hat\z}$.  Two types of uniaxial
anisotropies are introduced here: anisotropy in the exchange couplings
($J_z\neq J$) and single-ion anisotropy ($D\neq0$) which generally exist
owing to the crystal field and spin--orbit interaction.  Spin operators
$S_\r^\pm\equiv S_\r^x\pm iS_\r^y$ and $S_\r^z$ obey the usual
commutation relations $[S_\r^+,S_{\r'}^-]=2S_\r^z\delta_{\r,\r'}$,
$[S_\r^z,S_{\r'}^\pm]=\pm S_\r^\pm\delta_{\r,\r'}$ and the identity
$(S_\r^+)^{2S+1}=(S_\r^-)^{2S+1}=0$ for a spin $S$ representation.
$J,\,J_z>0$ corresponds to a ferromagnetic coupling and $J,\,J_z<0$ to
an antiferromagnetic coupling.  In the latter case, by rotating spins by
$\pi$ along the $z$-axis ($S_\r^\pm\to-S_\r^\pm$) only for sites with
odd-valued $x+y+z$, we can choose $J>0$, which is assumed from now on.

The ground state for a sufficiently large magnetic field $B<0$ is a
fully polarized state with all spins pointing downwards;
$S_\r^z|0\>=-S|0\>$ and $S_\r^-|0\>=0$.  Because of the U(1) symmetry of
the Hamiltonian (\ref{eq:hamiltonian}) under rotation
$S_\r^\pm\to e^{\pm i\theta}S_\r^\pm$, the relative magnetization
$N\equiv\sum_\r(S_\r^z+S)$ is a conserved quantity.  Accordingly, we can
consider an excited state with fixed $N$, which corresponds to a
particle number of magnons.  $N$-magnon excitations on this ground state
are described by a wave function
$\Psi(\r_1,\ldots,\r_N)\equiv\<0|\prod_{i=1}^NS_{\r_i}^-|\Psi\>$, which
is symmetric under any exchange of coordinates (Bose statistics) and
satisfies the Schr\"odinger equation:
\begin{align}\label{eq:schrodinger}
 & E\Psi(\r_1,\ldots,\r_N)
 = \<0|\Biggl[\prod_{i=1}^NS_{\r_i}^-,H\Biggr]|\Psi\> \notag\\
 &= \Biggl[\sum_{i=1}^N\sum_\e SJ\,(1-\nabla_\e^{^{}i})
 + \sum_{i<j}^N\Biggl\{\sum_\e J\,\delta_{\r_i,\r_j}\nabla_\e^{^{}i} \notag\\
 &\quad - \sum_\e J_z\,\delta_{\r_i,\r_j+\e}
 - 2D\delta_{\r_i,\r_j}\Biggr\}\Biggr]\Psi(\r_1,\ldots,\r_N).
\end{align}
Here $\nabla_\e^{^{}i}$ is an operator to displace $\r_i$ to $\r_i+\e$
and a constant energy shift $-(6SJ-6SJ_z-2SD+D+B)N$ is omitted in the
right hand side.

The emergence of the Efimov effect in quantum magnets can be understood
intuitively by using an exact mapping between spins and bosons, which is
known as the Holstein--Primakoff transformation~\cite{Holstein:1940}.
It is clear from the Schr\"odinger equation (\ref{eq:schrodinger}) that
$J$ in the first term acts as a hopping amplitude and gives a
single-magnon dispersion relation:
\begin{align}\label{eq:dispersion}
 E_0(\k) = \sum_\e SJ\,[1-\cos(\k\cdot\e)].
\end{align}
The rest describe interactions between a pair of magnons where $J_z>J$
and $D>0$ act as nearest-neighbour and on-site attractions,
respectively.\footnote[2]{Note that the effective $N$-magnon hardcore
repulsion also exists for $N=2S+1$ according to $(S_\r^+)^{2S+1}=0$.  By
setting $\r_1=\cdots=\r_N$ in equation (\ref{eq:schrodinger}), we obtain
$[E-3N(N-1)J+N(N-1)D]\Psi(\r,\ldots,\r)=0$.  Therefore, unless
$E-3N(N-1)J+N(N-1)D=0$, the constraint $\Psi(\r,\ldots,\r)=0$ is
automatically satisfied.  In particular, $D$ plays no role for $S=1/2$
as $\Psi(\r,\r)=0$.}  By tuning these couplings, we can induce a
scattering resonance between two magnons.  Once the two-magnon resonance
is achieved, three magnons should exhibit the Efimov effect
(\ref{eq:scaling}) because the Efimov effect is universal in the sense
that it emerges regardless of microscopic details.

\begin{table*}
 \caption{{\bf Few lowest binding energies of three magnons right at the
 two-magnon resonances.}  \label{tab:binding}}
 \begin{ruledtabular}
  \begin{tabular}{ccccccccccccc}
   & $S$ && $J_z/J$ && $D/J$ && $n$ && $E_n/J$ && $\sqrt{E_{n-1}/E_n}$ & \\[2.5pt]\hline\\[-12pt]
   & 1/2 && 2.93654 && ---  && 0 && $-2.09\times10^{-1}$ && --- & \\
   & && && && 1 && $-4.15\times10^{-4}$ && 22.4 & \\
   & && && && 2 && $-8.08\times10^{-7}$ && 22.7 & \\[1.5pt]\hline\\[-12pt]
   & 1 && 4.87307 && 0 && 0 && $-5.16\times10^{-1}$ && --- & \\
   & && && && 1 && $-1.02\times10^{-3}$ && 22.4 & \\
   & && && && 2 && $-2.00\times10^{-6}$ && 22.7 & \\[1.5pt]\hline\\[-12pt]
   & 1 && $+1$ && 4.76874 && 0 && $-5.50\times10^{-2}$ && --- & \\
   & && && && 1 && $-1.16\times10^{-4}$ && 21.8 & \\[1.5pt]\hline\\[-12pt]
   &  1  && $-1$ && 5.12703 && 0 && $-4.36\times10^{-3}$ && --- & \\
   & && && && 1 && $-8.88\times10^{-6}$ && 22.2 & \\[1.5pt]\hline\\[-12pt]
   & --- && --- && --- && $\infty$ && --- && \phantom{000}22.6944 & 
  \end{tabular}
 \end{ruledtabular}
\end{table*}

\subsection{Two-magnon resonance}
We start with a scattering problem of two magnons.  A two-magnon
solution with a center-of-mass momentum $\K$ is written as
\begin{align}
 \Psi(\r_1,\r_2) = e^{i\K\cdot\R}\,\psi_\K(\rh),
\end{align}
where $\R\equiv(\r_1+\r_2)/2$ and $\rh\equiv\r_1-\r_2$ are center-of-mass
and relative coordinates, respectively.  The Bose statistics of magnons
implies $\psi_\K(\rh)=\psi_\K(-\rh)$ for the relative wave function.
The two-magnon Schr\"odinger equation can be solved in a standard way by
bringing it into the Lippmann--Schwinger equation (see Methods for
details).

The scattering resonance between two magnons is defined by the
divergence of the $s$-wave scattering length ($a_s$) where a two-magnon
bound state appears from the continuum.  $a_s$ can be inferred from the
asymptotic behaviour of the wave function at zero energy and zero
center-of-mass momentum:
\begin{align}
 \lim_{|\rh|\to\infty}\psi_\0(\rh)\big|_{E=0}
 \propto 1 - \frac{a_s}{|\rh|}.
\end{align}
By matching this asymptotic behaviour with the obtained solution
(equation (\ref{eq:2-magnon_sol}) in Methods), the analytic expression
of $a_s$ is obtained as
\begin{align}\label{eq:scattering}
 \frac{a_s}{a} = \frac{\frac3{2\pi}\left[1-\frac{D}{3J}
 - \frac{J_z}{J}\left(1-\frac{D}{6SJ}\right)\right]}
 {2S-1 + \frac{J_z}{J}\left(1-\frac{D}{6SJ}\right)
 + 3_{}W\left[1-\frac{D}{3J}
 - \frac{J_z}{J}\left(1-\frac{D}{6SJ}\right)\right]}.
\end{align}
Here $a$ is the lattice constant and
$W\equiv\frac{\sqrt6}{96\pi^3}\,\Gamma\!\left(\frac1{24}\right)
\Gamma\!\left(\frac5{24}\right)\Gamma\!\left(\frac7{24}\right)
\Gamma\!\left(\frac{11}{24}\right)=0.505462$ results from one of the
Watson's triple integrals~\cite{Watson:1939}.

\begin{figure}
 \includegraphics[width=0.9\columnwidth,clip]{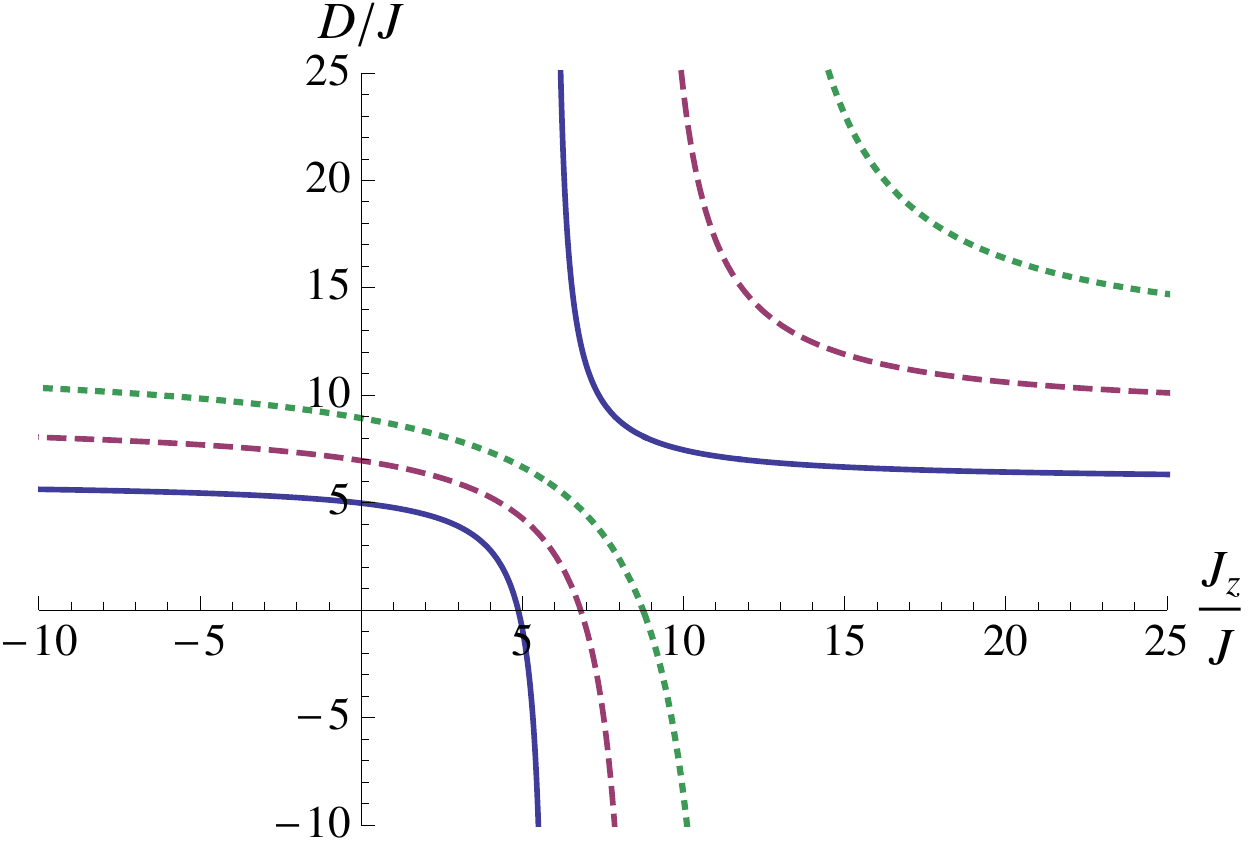}
 \caption{{\bf Critical anisotropy for the two-magnon resonance.}  The
 solid, dashed and dotted curves are for $S=1$, $S=3/2$ and $S=2$,
 respectively.  There exist(s) no, one or two $s$-wave bound state(s) at
 $\K=\0$ in the lower left, middle or upper right region, respectively,
 separated by the two critical curves.  The appearance of two degenerate
 $d$-wave bound states is not indicated here.  \label{fig:critical}}
\end{figure}

As a consequence, we find that the two-magnon resonance $a_s\to\infty$
takes place at $J_z/J=2.93654$ for $S=1/2$.  Critical anisotropies for
other spins are shown in Fig.~\ref{fig:critical}.  In particular, the
two-magnon resonance for $S=1$ is induced by the exchange anisotropy at
$J_z/J=4.87307$ ($D=0$), whereas it is induced by the single-ion
anisotropy at $D/J=4.76874$ ($J_z=J$) or $D/J=5.12703$ ($J_z=-J$) for an
isotropic ferromagnetic or antiferromagnetic coupling, respectively.
Above these critical points, two magnons form an $s$-wave bound state at
$\K=\0$.  Its binding energy $E<0$ determines the three-magnon threshold
in Figs.~\ref{fig:s_half} and \ref{fig:s_one} below.

\subsection{Three-magnon Efimov effect}
We now turn to a bound-state problem of three magnons.  A bound-state
solution to the three-magnon Schr\"odinger equation can be obtained in a
similar way to the previous two-magnon problem (see Methods for
details).  Because our purpose here is to demonstrate the Efimov effect
of magnons, we focus on the $s$-wave channel at zero center-of-mass
momentum where the Efimov effect is supposed to emerge.

We find that three magnons form a series of bound states and their
binding energies are shown in Table~\ref{tab:binding} right at the
anisotropy-induced two-magnon resonances located above.  The ratios of
two successive binding energies obey the universal scaling law
(\ref{eq:scaling}), which supports that the observed bound states of
three magnons are indeed the Efimov states.  Note that although the
Efimov effect emerges only in the $s$-wave
channel~\cite{Nielsen:2001,Braaten:2006}, it is in general possible that
there are non-Efimov three-magnon bound states in other channels, which
may have lower binding energies.

\begin{figure}
 \includegraphics[width=0.96\columnwidth,clip]{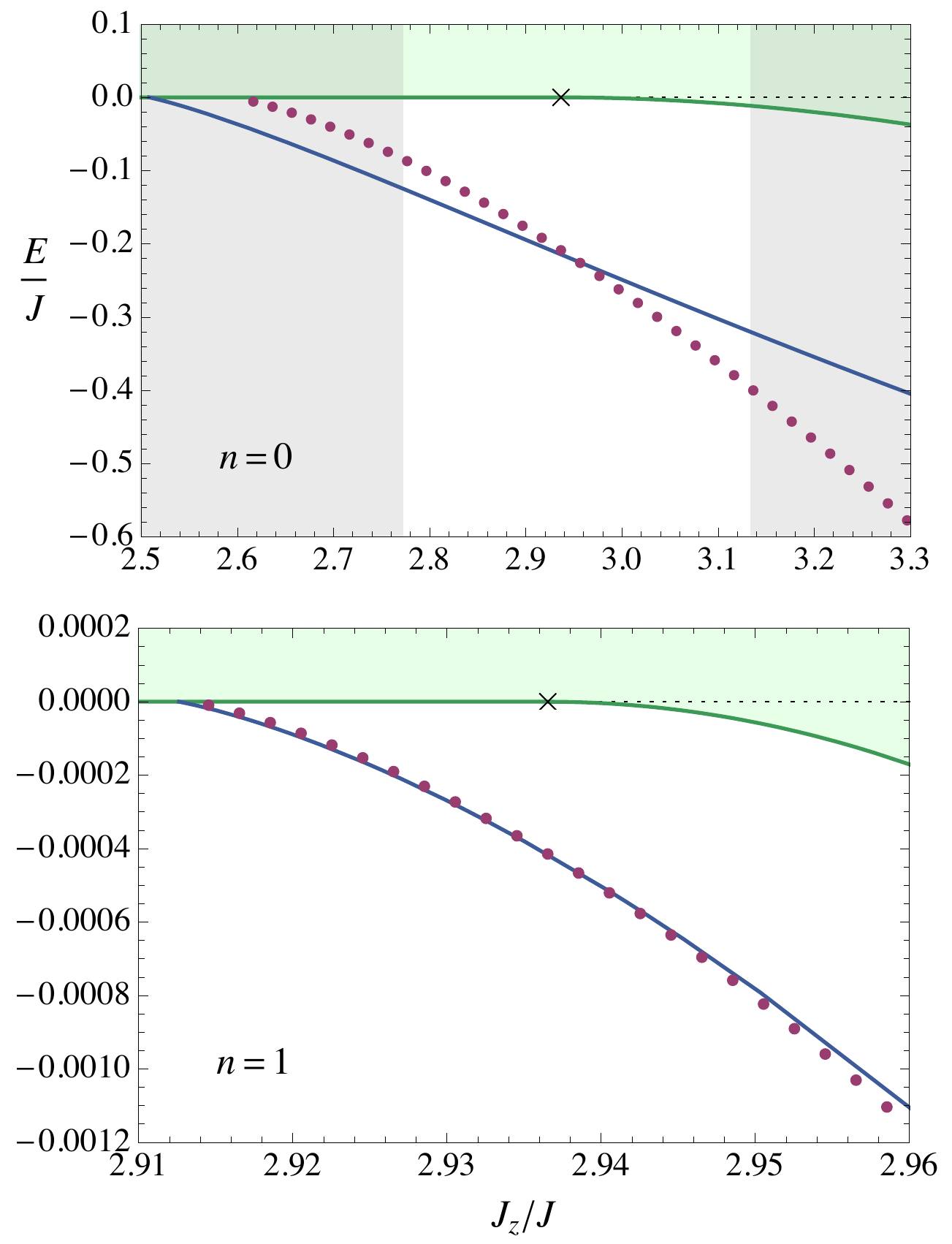}
 \caption{{\bf Two lowest binding energies of three magnons for
 $\bm{S=1/2}$ near the two-magnon resonance.}  Dots are numerical
 results and the solid curve is from the universal theory
 (\ref{eq:universal}).  The upper coloured region represents the
 three-magnon continuum and the cross ($\times$) marks the critical
 point where a two-magnon bound state appears.  The left and right gray
 areas for $n=0$ indicate regions where $|a_s|/a<10$.
 \label{fig:s_half}}
\end{figure}

In the case of $S=1/2$, the two lowest binding energies near the
two-magnon resonance are shown in Fig.~\ref{fig:s_half}.  The universal
theory predicts that the spectrum of Efimov states is completely
characterized by the $s$-wave scattering length $a_s$ and the so-called
Efimov parameter $\kappa_*$~\cite{Braaten:2006}:
\begin{align}\label{eq:universal}
 E_n \to -\lambda^{-2n}\,\frac{\hbar^2\kappa_*^2}{m}\,
 F\!\left(\frac{\lambda^n}{\kappa_*a_s}\right) \qquad (n\to\infty).
\end{align}
Here $m$ is the mass of constituent particles and $F(x)$ is the
universal function defined in a range $-0.663293\leq x\leq14.1314$ and
normalized as $F(0)=1$.  The inverse effective mass of magnons is
$1/m=2SJa^2/\hbar^2$, which is inferred from the single-magnon
dispersion relation (\ref{eq:dispersion}).  By matching equation
(\ref{eq:universal}) with the binding energy of the second excited state
at the resonance, we obtain $\kappa_*a\simeq0.463$.  The resulting
universal curves for $n=0$ and $n=1$ are also plotted in
Fig.~\ref{fig:s_half} by using $a_s$ obtained in equation
(\ref{eq:scattering}).  We find an excellent agreement of the binding
energy of the first excited state with the universal theory, which
leaves no doubt that this three-magnon bound state is the universal
Efimov state.  On the other hand, the binding energy of the ground state
deviates from the universal theory away from the two-magnon resonance
where non-universal corrections become non-negligible.

\begin{figure}
 \includegraphics[width=0.96\columnwidth,clip]{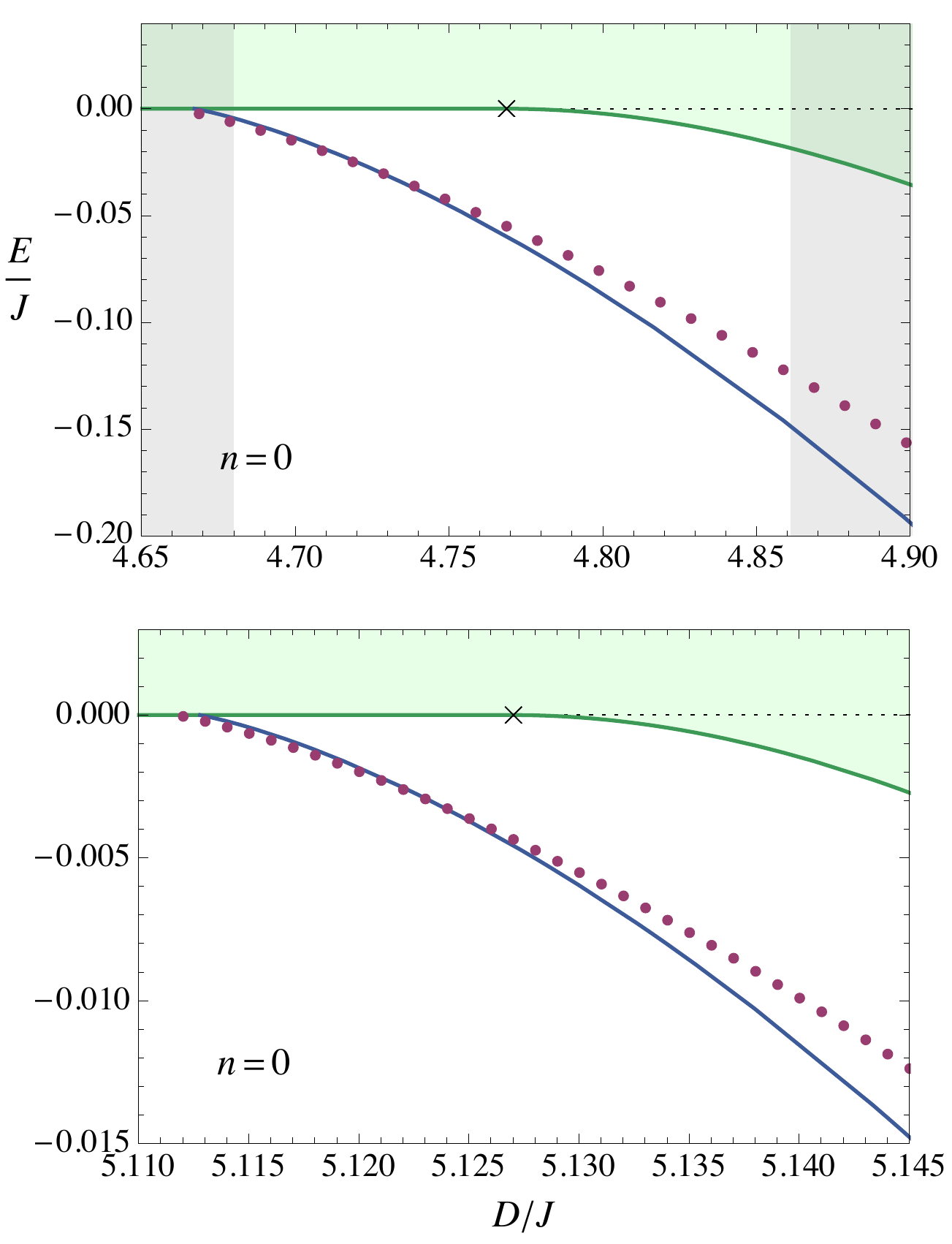}
 \caption{{\bf Lowest binding energy of three magnons for $\bm{S=1}$
 near the two-magnon resonance.}  The upper (lower) panel corresponds to
 $J_z=J$ ($-J$).  See the legend of Fig.~\ref{fig:s_half} for other
 details.  \label{fig:s_one}}
\end{figure}

Similarly, in the case of $S=1$ with $J_z=J$ ($-J$), we obtain
$\kappa_*a\simeq0.173$ ($0.0478$) by matching equation
(\ref{eq:universal}) with the binding energy of the first excited state
at the resonance.  The resulting universal curve for $n=0$ is plotted in
Fig.~\ref{fig:s_one} as well as the ground-state binding energy near the
two-magnon resonance.  We find a reasonable agreement between them,
which confirms that this three-magnon bound state is consistent with the
universal Efimov state.  Our findings here support the fact that
resonantly interacting magnons fall into the class of universal few-body
systems.  Accordingly, other universal aspects of Efimov physics, such
as a pair of four-body resonances associated with every Efimov
state~\cite{Hammer:2007,Stecher:2009}, also apply to the system of
magnons.

\subsection{Towards experimental realization}
In summary, we showed that magnons in quantum magnets exhibit the Efimov
effect by tuning an easy-axis exchange or single-ion anisotropy.  The
single-ion anisotropy $D$ can be changed significantly in organic
magnets by choosing different ligands~\cite{Blundell:2004}.  Therefore,
it is possible to find a compound with spin $S\geq1$ whose $D/J$ ratio
is already close to the critical value.  For example, there is an $S=1$
ferromagnetic compound based on molecular Ni$^{2+}$ squares with
$D/J\simeq3.0(5)$~\cite{Koch:2003}, which is not far from the critical
value $4.76874$.  Furthermore, the exchange coupling $J$ can be tuned
with pressure~\cite{Kawamoto:2001} to bring the system near the
two-magnon resonance and realize the Efimov effect.

\begin{figure}
 \includegraphics[width=0.9\columnwidth,clip]{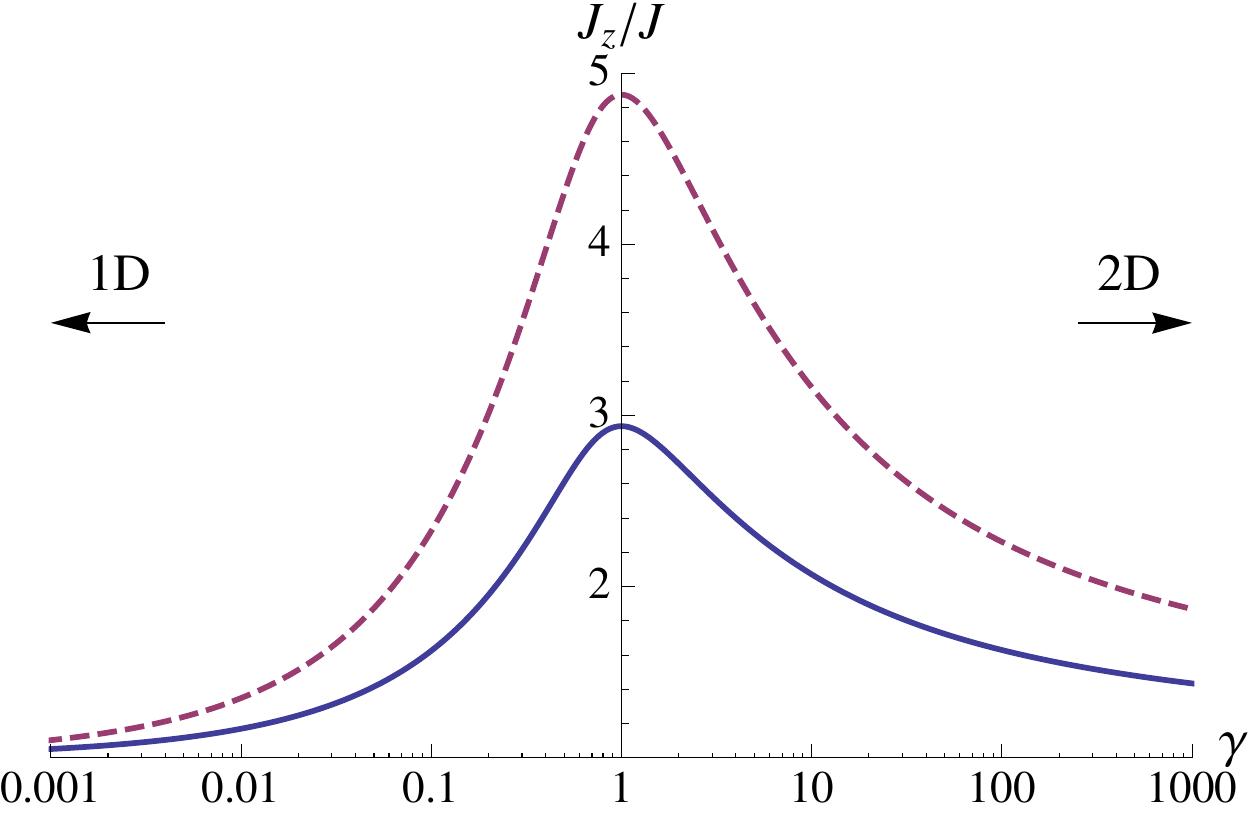}
 \caption{{\bf Critical exchange anisotropy for the two-magnon resonance
 in the presence of the spatial anisotropy.}  The solid and dashed
 curves are for $S=1/2$ and $S=1$ with $D=0$, respectively.  An $s$-wave
 bound state at $\K=\0$ appears in the upper side of the critical curve.
 \label{fig:anisotropy}}
\end{figure}

On the other hand, the single-ion anisotropy plays no role for $S=1/2$.
In this case, the two-magnon resonance can be induced by the exchange
anisotropy, although it is in general difficult to tune and its critical
value ($J_z/J=2.93654$) is somewhat large.  However, the critical
exchange anisotropy can be reduced significantly if the magnet is
spatially anisotropic.  Spatially anisotropic exchange couplings can be
taken into account simply by replacing $J$ and $J_z$ in the Hamiltonian
(\ref{eq:hamiltonian}) with $J^\e$ and $J_z^\e$, respectively.  We
assume a uniaxial anisotropy $J\equiv J^{\hat\x}=J^{\hat\y}$ and
$J_z\equiv J_z^{\hat\x}=J_z^{\hat\y}$ with a shared ratio
$\gamma\equiv J/J^{\hat\z}=J_z/J_z^{\hat\z}$.  As shown in
Fig.~\ref{fig:anisotropy}, the corresponding critical exchange
anisotropy reduces significantly towards an isotropic point $J_z/J\to1$
in a quasi-one-dimensional ($\gamma\to0$) or two-dimensional
($\gamma\to\infty$) limit.  Because magnets with strong spatial
anisotropies are very
common~\cite{Landee:1979,Takahashi:1991,Feldkemper:1995,Manaka:2003,Shimizu:2006,Sugano:2010,Pradeep:1999},
there is hope to find an $S=1/2$ ferromagnetic compound whose $J_z/J$
ratio is already close to the critical value.  Once such a compound is
identified, the spatial anisotropy $\gamma$ can be tuned with pressure.

\begin{figure}
 \includegraphics[width=0.92\columnwidth,clip]{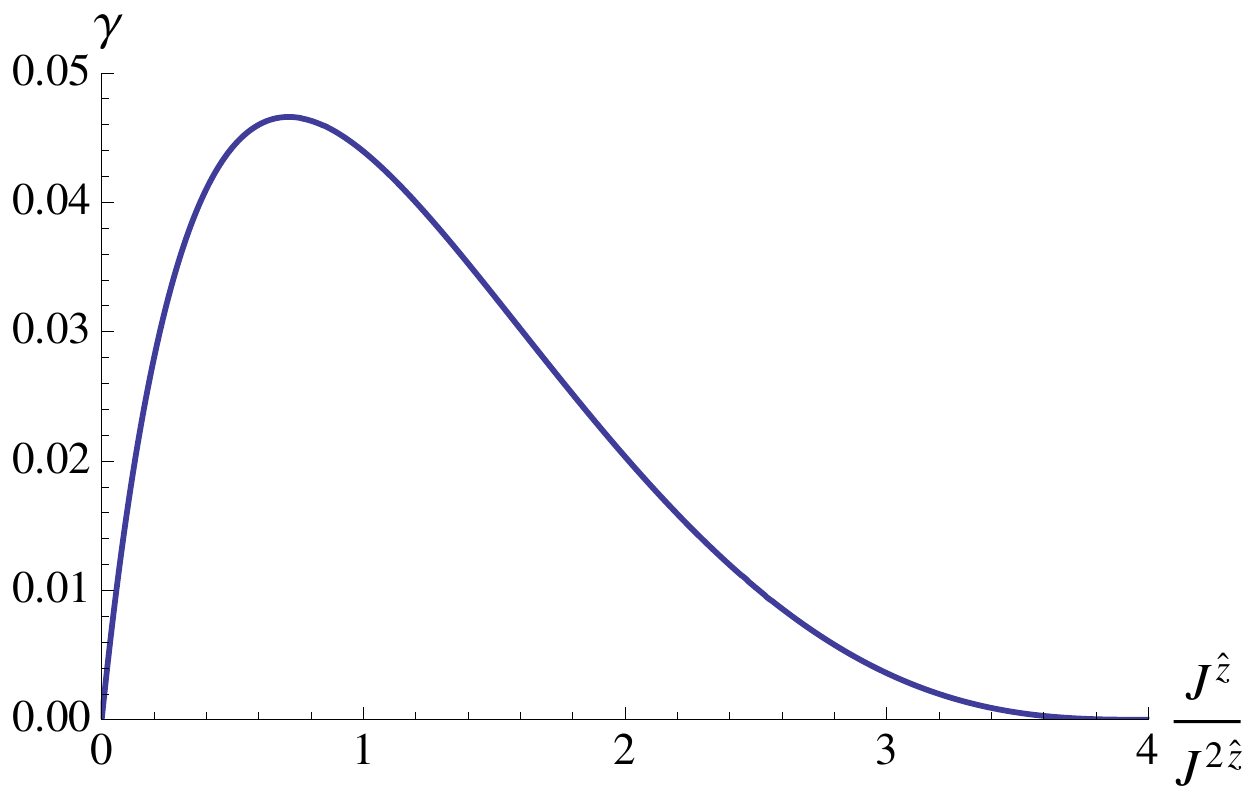}
 \caption{{\bf Critical spatial anisotropy for the two-magnon resonance
 for $\bm{S=1/2}$ in the presence of the frustration.}  An $s$-wave
 bound state at $\K=(0,0,2k_0)$ appears in the lower side of the
 critical curve.  The critical spatial anisotropy for $S=1$ with $D=0$
 is at most $\gamma<3.61\times10^{-5}$ and thus invisible here.
 \label{fig:frustration}}
\end{figure}

An alternative approach to induce the two-magnon resonance in $S=1/2$
magnets even without the exchange anisotropy ($J_z=J$) is to introduce
frustrated exchange interactions.  The simplest example is given by
quasi-one-dimensional spin chains with nearest-neighbour ferromagnetic
and next-nearest-neighbour antiferromagnetic couplings:
\begin{align}
 H_\mathrm{intra} = -\sum_\r(J^{\hat\z}\S_\r\cdot\S_{\r+\hat\z}
 - J^{2\hat\z}\S_\r\cdot\S_{\r+2\hat\z}),
\end{align}
which are realized in Rb$_2$Cu$_2$Mo$_3$O$_{12}$~\cite{Hase:2004},
LiCuVO$_4$~\cite{Enderle:2005} and Li$_2$CuO$_2$~\cite{Dmitriev:2009}.
When $J^{\hat\z}/J^{2\hat\z}<4$, the single-magnon dispersion develops
minima at nonzero momentum
$k_0\equiv\pm\arccos(J^{\hat\z}/4J^{2\hat\z})$, and accordingly, two
magnons form a bound state with a center-of-mass momentum
$2k_0$~\cite{Bahurmuz:1986,Chubukov:1991,Cabra:2000,Dmitriev:2006,Kuzian:2007,Kecke:2007,Dmitriev:2009}.
This two-magnon bound state disappears into the continuum at a certain
interchain coupling:
\begin{align}
 H_\mathrm{inter} = -\sum_\r J\,(\S_\r\cdot\S_{\r+\hat\x}
 + \S_\r\cdot\S_{\r+\hat\y}),
\end{align}
which leads to the two-magnon resonance.  The corresponding critical
spatial anisotropy $\gamma=J/J^{\hat\z}$ is shown in
Fig.~\ref{fig:frustration}.  Because the frustration ratio
$J^{\hat\z}/J^{2\hat\z}$ is highly tunable with pressure owing to the
strong dependence of ferromagnetic couplings on the
cation--anion--cation angle of the superexchange
path~\cite{Goodenough:1963}, $S=1/2$ frustrated magnets are promising
candidates for realizing the Efimov effect without the strong exchange
anisotropy.  The same approach can also be applied to
quasi-two-dimensional frustrated magnets~\cite{Nath:2008}.

So far multi-magnon bound states have been observed mostly in
quasi-one-dimensional compounds but with different experimental
techniques, such as absorption
spectroscopy~\cite{Date:1966,Date:1968,Torrance:1969th,Torrance:1969ex,Nicoli:1974,Vlimmeren:1979,Vlimmeren:1980,Bosch:1987,Perkins:1995},
inelastic neutron scattering~\cite{Garrett:1997,Tennant:2003} and
electron spin
resonance~\cite{Hoogerbeets:1984,Nijhof:1986,Rubins:1994,Orendac:1999,Katsumata:2000,Zvyagin:2007,Psaroudaki:2012}.
The same spectroscopic measurements can be used here to observe the
emergent Efimov states of magnons, provided that the conservation of the
magnetization along the magnetic field axis is weakly
violated~\cite{Torrance:1969th,Torrance:1969ex,Katsumata:2000}.  We
found that even the lowest bound state of three magnons is already
consistent with the universal Efimov state.  Its binding energy for
ferromagnetic cases is $5\sim55\%$ of the exchange coupling, which can
be up to $10^2\sim10^3$~K.  As a dilution refrigerator can lower the
temperature down to a few tens of millikelvin, the observation of the
lowest one or two Efimov state(s) is within reach.

\subsection{Methods}

\subsubsection{Two-magnon problem}
The Schr\"odinger equation (\ref{eq:schrodinger}) for two magnons
($N=2$) can be solved in a standard way.  By treating the first term in
the right hand side as a free part ($H_0$) and the rest as an
interaction part ($V$), the two-magnon Schr\"odinger equation can be
brought into the Lippmann--Schwinger equation:
\begin{align}\label{eq:2-magnon_sol}
 & \psi_\K(\rh) = \phi_\K(\rh) + \<\rh|\frac1{E-H_0+i0^+}V|\psi_\K\> \notag\\
 &= \phi_\K(\rh) + \int_{-\pi/a}^{\pi/a}\!\frac{d\k}{(2\pi/a)^3}
 \frac{\cos(\k\cdot\rh)}{E-E_\K(\k)+i0^+} \notag\\
 &\times \left[\sum_\e\{J\cos(\tfrac\K2\cdot\e)
 - J_z\cos(\k\cdot\e)\}\psi_\K(\e) - 2D\psi_\K(\0)\right].
\end{align}
Here $\phi_\K(\rh)$ is a solution to $E\phi_\K(\rh)=H_0\phi_\K(\rh)$ and
$E_\K(\k)\equiv E_0(\frac\K2+\k)+E_0(\frac\K2-\k)$ is the energy of two
non-interacting magnons.  By setting $\rh=\hat\x,\,\hat\y,\,\hat\z$ and
$\rh=\0$ in equation (\ref{eq:2-magnon_sol}), we obtain four coupled
equations which determine the four unknown constants; $\psi_\K(\e)$ and
$\psi_\K(\0)$, which in turn determine $\psi_\K(\rh)$ through equation
(\ref{eq:2-magnon_sol}).  The same result was obtained in the isotropic
case of $J_z=J$ and $D=0$ in the pioneering work~\cite{Wortis:1963}.

\subsubsection{Three-magnon problem}
The Schr\"odinger equation (\ref{eq:schrodinger}) for three magnons
($N=3$) can be solved in a similar way.  By treating the first term in
the right hand side as a free part ($H_0$) and the rest as an
interaction part ($V$), the Lippmann--Schwinger equation for $E<0$ is
written as
\begin{align}\label{eq:3-magnon_LS}
 \Psi(\r_1,\r_2,\r_3) = \<\r_1,\r_2,\r_3|\frac1{E-H_0}V|\Psi\>.
\end{align}
We then introduce a new parametrization of the wave function:
\begin{align}
 \chi_\K(\rh;\k) &\equiv \sum_{\r_1,\r_2,\r_3}\delta_{\rh,\r_1-\r_2}\,
 e^{-i(\K-\k)\cdot\left(\frac{\r_1+\r_2}2\right)-i\k\cdot\r_3} \notag\\
 &\quad \times \Psi(\r_1,\r_2,\r_3),
\end{align}
which describes three magnons with a center-of-mass momentum $\K$
in which two of them are separated by a distance $\rh$ and the third one
has a momentum $\k$.  The Bose statistics of magnons
implies $\chi_\K(\rh;\k)=\chi_\K(-\rh;\k)$.

After a straightforward calculation, the Lippmann--Schwinger equation
(\ref{eq:3-magnon_LS}) can be brought into
\begin{align}\label{eq:3-magnon_sol}
 &\hspace{-2pt} \chi_\K(\rh;\k_3) = \int_{-\pi/a}^{\pi/a}\!\frac{d\k_1}{(2\pi/a)^3}
 \frac{\cos(\frac{2\k_1+\k_3-\K}2\cdot\rh)}{E-E_0(\k_1,\K-\k_1-\k_3,\k_3)} \notag\\
 &\hspace{-2pt}\times \Biggl[\sum_\e\{J\cos(\tfrac{\K-\k_3}2\cdot\e)
 -J_z\cos(\tfrac{2\k_1+\k_3-\K}2\cdot\e)\}\chi_\K(\e;\k_3) \notag\\
 &\hspace{-2pt}\quad + 2\sum_\e\{J\cos(\tfrac{\K-\k_1}2\cdot\e)
 -J_z\cos(\tfrac{\k_1+2\k_3-\K}2\cdot\e)\} \notag\\[-4pt]
 &\hspace{-2pt}\quad \times \chi_\K(\e;\k_1)
 - 2D\chi_\K(\0;\k_3) - 4D\chi_\K(\0;\k_1)\Biggr],
\end{align}
where $E_0(\k_1,\k_2,\k_3)\equiv\sum_{i=1,2,3}E_0(\k_i)$ is the energy
of three non-interacting magnons.  By setting
$\rh=\hat\x,\,\hat\y,\,\hat\z$ and $\rh=\0$ in equation
(\ref{eq:3-magnon_sol}), we obtain four coupled integral equations which
determine the allowed binding energy $E<0$ and the four unknown
functions; $\chi_\K(\e;\k)$ and $\chi_\K(\0;\k)$, which in turn
determine $\chi_\K(\rh;\k)$ through equation (\ref{eq:3-magnon_sol}).

Bound-state solutions in the $s$-wave channel at $\K=\0$ correspond to
those where (i)
$\chi_\0(\hat\x;k_x,k_y,k_z)=\chi_\0(\hat\y;k_z,k_x,k_y)=\chi_\0(\hat\z;k_y,k_z,k_x)$,
(ii) $\chi_\0(\hat\x;k_x,k_y,k_z)=\chi_\0(\hat\x;k_x,k_z,k_y)$, (iii)
$\chi_\0(\0;k_x,k_y,k_z)$ is symmetric under any exchange among
$\{k_x,k_y,k_z\}$ and (iv) all of them are even functions of $k_\nu$
($\nu=x,y,z$).  The resulting two coupled integral equations with three
variables ranging from $0$ to $\pi$ are solved numerically by
discretizing each variable with the Gaussian quadrature rule.

\subsection{Acknowledgements}
This work was supported by a LANL Oppenheimer Fellowship and the US DOE
contract No.\ DE-AC52-06NA25396 through the LDRD program.  All authors
contributed equally to this work and declare no competing financial
interests.  Correspondence and requests for materials should be
addressed to Y.\,N.

\end{document}